\documentclass[12pt]{article}
\usepackage{epsfig}
\usepackage{colortbl}
\usepackage{graphics}
\usepackage{graphicx}
\def\be{\begin{equation}}
\def\ee{\end{equation}}
\def\bea{\begin{eqnarray}}
\def\eea{\end{eqnarray}}
\usepackage{graphicx}

\definecolor{red}{rgb}{1.00,0.00,0.00}

\catcode`\@=11
\def\lsim{\mathrel{\mathpalette\@versim<}}
\def\gsim{\mathrel{\mathpalette\@versim>}}
\def\@versim#1#2{\vcenter{\offinterlineskip
\ialign{$\m@th#1\hfil##\hfil$\crcr#2\crcr\sim\crcr } }}
\catcode`\@=12

\catcode`@=12
\begin{document}
\thispagestyle{empty}
\begin{flushright}
UCRHEP-T526\\
March 2013\
\end{flushright}
\vspace{0.6in}
\begin{center}
{\LARGE \bf Updated $S_3$ Model of Quarks\\}
\vspace{1.2in}
{\bf Ernest Ma$^1$ and Bla\v zenka Meli\'c$^{1,2}$\\}
\vspace{0.2in}
{\sl $^1$ Department of Physics and Astronomy, University of
California,\\
Riverside, California 92521, USA\\}
\vspace{0.1in}
{\sl $^2$ Theoretical Physics Division, Rudjer Bo\v skovi\'c Institute,\\ 
10002 Zagreb, Croatia\\}
\end{center}
\vspace{0.6in}
\begin{abstract}\
A model proposed in 2004 using the non-Abelian discrete symmetry $S_3$ 
for understanding the flavor structure of quarks and leptons is updated, 
with special focus on the quark and scalar sectors.  We show how the 
approximate residual symmetries of this model explain both the pattern 
of the quark mixing matrix and why the recently observed particle of 
126 GeV at the Large Hadron Collider is so much like the one Higgs 
boson of the Standard Model.  We identify the strongest phenomenological 
bounds on the scalar masses of this model, and predict a possibly 
observable decay $b \to s \tau^- \mu^+$, but not $b \to s \tau^+ \mu^-$.
\end{abstract}

\newpage
\baselineskip 24pt

\section{Introduction}

With the discovery~\cite{atlas12,cms12} of a particle of 126 GeV at the Large 
Hadron Collider (LHC) and no evidence for any others in a wide range of 
masses, extensions of the standard model are now severely constrained. 
In particular, if we want to understand the pattern of quark and lepton 
masses and their mixing in terms of a flavor symmetry, we are faced with 
a new theoretical challenge.  In order to carry the flavor symmetry in 
the context of a renormalizable theory, new scalar multiplets are 
required.  We now must have a good reason within the flavor model as 
to why the one observed light Higgs boson is so much like that of the 
Standard Model (SM).  Of course we need also to understand within the same 
context of why the quark mixing matrix is nearly diagonal, whereas 
the neutrino (lepton) mixing matrix is close to tribimaximal.

In 2004, the family structure of quarks and leptons was explained in a 
model~\cite{cfm04} using the non-Abelian discrete symmetry 
$S_3$.  It has a symmetry breaking pattern designed to allow for small 
$2-3$ mixing in the quark sector and near-maximal $2-3$ mixing in the neutrino 
sector.  Whereas all the basic details were described for both quarks 
and leptons, that paper dealt mainly with neutrino mixing, with the 
specific assumption of negligible $e-\mu$ mixing in the charged-lepton mass 
matrix, although this $1-2$ mixing is generally allowed by the $S_3$ symmetry 
and is unavoidable also in the quark sector. 
As a result of that arbitrary assumption, the neutrino mixing angle 
$\theta_{13}$ was predicted to be very small: $0.02 \pm 0.01$.  Given that 
it has now been measured~\cite{daya12,reno12} at about 0.16, this prediction 
based on that arbitrary assumption is certainly ruled out, and the observed 
value of $\theta_{13}$ should be attributed to $e-\mu$ mixing 
within the context of this model.  The Higgs sector of this leptonic model 
has also been studied~\cite{blp11,blp12} for its collider signatures. 

In this paper we study the quark sector itself in detail and show how the 
Higgs sector is constrained by present data.  In particular, we identify 
two approximate residual discrete $Z_2$ symmetries, which allow the 
lightest scalar particle to be the observed 126 GeV particle, with the 
property that it is naturally very close to that of the SM. 

In Sec.~2 we present the $S_3$ model of Ref.~\cite{cfm04}, writing down 
specifically all the quark and Higgs representations.  In Sec.~3 we 
discuss the $c-t$ and $s-b$ quark sectors and show how they align in a 
symmetry limit, and the resulting phenomenological constraint from 
these two sectors.  In Sec.~4 we add the $u$ and $d$ quarks and discuss the 
full $3 \times 3$ quark mixing matrix.  In Sec.~5 we consider the full scalar 
sector consisting of three Higgs doublets, and show how the one light neutral 
scalar of this sector resembles that of the standard model to a very good 
approximation, not being a result of fine tuning but based on symmetry. 
In Sec.~6 we derive the phenomenological constraint on the third Higgs 
doublet which is responsible for mixing the first family with the other two. 
In Sec.~7 we make a specific verifiable prediction of this model, i.e. 
$b \to s \tau^- \mu^+$ could have a branching fraction as large as 
$10^{-7}$, whereas $b \to s \tau^+ \mu^-$ would be suppressed by a 
relative factor of $m_\mu^2/m_\tau^2$.  In Sec.~8 we conclude.

\section{The $S_3$ Model}

The smallest non-Abelian discrete symmetry is the group $S_3$ of the 
permutation of three objects.  It has six elements, and is isomorphic 
to the symmetry group of the equilateral triangle (identity, rotations by 
$\pm 2\pi/3$, and three reflections).  It has three irreducible 
representations $\underline{1}, \underline{1}', \underline{2}$, with the 
multiplication rules:
\begin{eqnarray}
&& \underline{1} \times \underline{1}' = \underline{1}', ~~~  
\underline{1}' \times \underline{1}' = \underline{1}, ~~~ 
\underline{2} \times \underline{1} = \underline{2}, ~~~ 
\underline{2} \times \underline{1}' = \underline{2}, \\ 
&& \underline{2} \times \underline{2} = \underline{1} + \underline{1}' 
+ \underline{2} \,.
\end{eqnarray}
The specific choice of $2 \times 2$ matrices for the \underline{2} 
representation is only unique up to a unitary transformation.  A particular 
practical and elegant one appeared in Ref.~\cite{dgp92} which was followed 
in Ref.~\cite{cfm04}.  For a review, see Ref.~\cite{m04}.  In this 
representation, if
\begin{equation}
\pmatrix{a_1 \cr a_2}, ~ \pmatrix{b_1 \cr b_2} \sim \underline{2} \,,
\end{equation}
under $S_3$, then 
\begin{equation}
\pmatrix{a_2^\dagger \cr a_1^\dagger}, ~ \pmatrix{b_2^\dagger \cr b_1^\dagger} 
\sim \underline{2}\, ;
\end{equation}
so that
\begin{equation}
a_1 b_2 + a_2 b_1 \sim \underline{1}\,, ~~~ a_1 b_2 - a_2 b_1 \sim \underline{1}'\,, 
~~~ \pmatrix{a_2 b_2 \cr a_1 b_1} \sim \underline{2}\,,
\end{equation}
and
\begin{equation}
a_1^\dagger b_1 + a_2^\dagger b_2 \sim \underline{1}\,, ~~~ a^\dagger_1 b_1 - 
a_2^\dagger b_2 \sim \underline{1}'\,, ~~~ \pmatrix{a_1^\dagger b_2 \cr a_2^\dagger 
b_1} \sim \underline{2}\,.
\end{equation}
The consequence of this representation is the elegant result that the 
trilinear combination $a_1 b_1 c_1 + a_2 b_2 c_2$ is a singlet.

Consider all quarks as left-handed fields, so that the usual right-handed 
ones are represented by charge conjugates.  Let $Q_i = (u_i,d_i)$, then
we assign as in Ref.~\cite{cfm04}
\begin{equation}
Q_1, u^c, d^c, c^c, s^c \sim \underline{1}\,, ~~~ t^c, b^c \sim \underline{1}'\,, 
~~~ \pmatrix{Q_2 \cr Q_3} \sim \underline{2}\,.
\end{equation}
In analogy to the three quark families, there are also three Higgs doublets 
$\Phi_i = (\phi_i^0, \phi_i^-)$ with assignments:
\begin{equation}
\pmatrix{\Phi_1 \cr \Phi_2} \sim \underline{2}\,, ~~~ \Phi_3 \sim \underline{1}\,. 
\end{equation}

\section{The $c-t$ and $s-b$ Quark Sectors}

In this sector, only $\Phi_{1,2}$ are involved in the Yukawa interactions, 
i.e.
\begin{eqnarray}
-{\cal L}_Y &=& g_1^d [(\phi_1^0 b + \phi_2^0 s) - (\phi_1^- t + \phi_2^- c)] s^c 
+ g_2^d [(\phi_1^0 b - \phi_2^0 s) - (\phi_1^- t - \phi_2^- c)] b^c \nonumber \\  
&+& g_1^u [(\bar{\phi}_2^0 t + \bar{\phi}_1^0 c) + (\phi_2^+ b + \phi_1^+ s)] c^c 
+ g_2^u [(\bar{\phi}_2^0 t - \bar{\phi}_1^0 c) + (\phi_2^+ b - \phi_1^+ s)] t^c \,.
\end{eqnarray} 
Consider now the Higgs potential of $\Phi_{1,2}$.   In addition to the $S_3$ 
symmetrical term, we add a soft term which breaks $S_3$, but preserves the discrete 
$Z_2$ symmetry $\Phi_1 \leftrightarrow \Phi_2$.  Hence,
\begin{eqnarray}
V_{12} &=& \mu_1^2 (\Phi_1^\dagger \Phi_1 + \Phi_2^\dagger \Phi_2) 
- \mu_2^2 (\Phi_1^\dagger \Phi_2 + \Phi_2^\dagger \Phi_1) \nonumber \\ 
&+& {1 \over 2} \lambda_1 (\Phi_1^\dagger \Phi_1 + \Phi_2^\dagger \Phi_2)^2 
+ {1 \over 2} \lambda_2 (\Phi_1^\dagger \Phi_1 - \Phi_2^\dagger \Phi_2)^2 
+ \lambda_3 (\Phi_1^\dagger \Phi_2) (\Phi_2^\dagger \Phi_1)\,.
\end{eqnarray}
This has a minimum with $v_1=v_2=v=123$ GeV, where
\begin{equation}
\mu_1^2 - \mu_2^2 + (2 \lambda_1 + \lambda_3) v^2 = 0\,.
\end{equation}
As a result, the normalized physical scalar bosons and their masses 
are given by
\begin{eqnarray}
&& h^0 = {\rm Re}(\phi_{1} + \phi_{2})\,, ~~~ m^2(h^0) = 2(2\lambda_1 + \lambda_3)v^2\,, \\   
&& H^0 = {\rm Re}(\phi_{1} - \phi_{2})\,, ~~~ m^2(H^0) = 2 \mu_2^2 
+ 2(2\lambda_2 -\lambda_3)v^2\,, \\ 
&& A = {\rm Im}(\phi_{1} - \phi_{2}), ~~~ m^2(A) = 2 \mu_2^2\,, \\ 
&& H^\pm = {1 \over \sqrt{2}} (\phi_1^\pm - \phi_2^\pm)\,, ~~~ m^2(H^\pm) = 
2 \mu_2^2 - 2 \lambda_3 v^2\,.
\end{eqnarray} 
Note that without the $\mu_2^2$ term which breaks $S_3$, $A$ would be massless.
 
Consider now the generic structure of the $c-t$ and $s-b$  
mass matrices.  They are of the form given by Eq.~(12) in  Ref.~\cite{cfm04}, 
i.e.
\begin{equation}
{\cal M} = \pmatrix{g_1 v & -g_2 v \cr g_1 v & g_2 v} = 
{1 \over \sqrt{2}} \pmatrix{1 & -1 \cr 1 & 1} \pmatrix{g_1 \sqrt{2} v & 0 
\cr 0 & g_2 \sqrt{2} v}\,.
\end{equation}
Since they are both diagonalized by the same $2 \times 2$ unitary matrix, 
there is no mismatch, and the quark mixing matrix is diagonal, explaining 
to a good first approximation what is observed.  Note that this result 
is based on the symmetry breaking pattern $S_3 \to Z_2$.   Now $h^0$ may be 
identified with the corresponding SM Higgs boson which couples to 
$(m_s/{2}v) \bar{s} s$ and $(m_b/{2}v) \bar{b} b$, etc. 

As for $H^\pm,H^0,A$, their Yukawa couplings are given by
\begin{eqnarray}
{\cal L}_Y &=& {m_s \over \sqrt{2} v} \left[H^+ \bar{t}_L + 
\left( {H^0 + i A \over \sqrt{2}} \right) \bar{b}_L \right] s_R \nonumber \\ 
&+& {m_b \over \sqrt{2} v} \left[ H^+ \bar{c}_L + \left( {H^0 + i A \over 
\sqrt{2}} \right) \bar{s}_L \right] b_R + h.c.,~~{\rm etc.}
\end{eqnarray}
Hence $H^0$ and $A$ will contribute significantly to $B_s - \bar{B}_s$ 
mixing, with resulting bounds on their masses.  The tree-level effective 
four-fermion interaction is given by 
\begin{equation}
{m_b^2 \over 4v^2} \left( {1 \over m_H^2} - {1 \over m_A^2} \right) 
(\bar{s}_L b_R)^2 + {m_s m_b \over 4v^2} \left( {1 \over m_H^2} + 
{1 \over m_A^2} \right) (\bar{s}_L b_R)(\bar{s}_R b_L)\,. 
\end{equation}
It adds to the usual SM box-diagram contribution for $B_s - \bar{B}_s$ 
mixing. The two new effective four-quark operators are usually denoted 
by ${\cal O}_2$  and ${\cal O}_4$~\cite{buras1}, i.e.
\begin{eqnarray}
{\cal O}_2 &=& (\overline{s}_L b_R)^2 \,, 
\\
{\cal O}_4 &=& (\overline{s}_L b_R) (\overline{s}_R b_L) \,. 
\end{eqnarray}
The matrix elements of the operators are calculated at the $m_{H,A}$ mass 
scale and evolved to the hadronic scale by using the anomalous dimension 
matrices given in Ref.~\cite{ciuchini2,buras3,buras2}. 
The mass difference in the $B_s- \bar{B}_s$ system is then given by 
\begin{eqnarray}
\Delta M_s = (\Delta M_s)_{\rm SM} + (\Delta M_s)_{{\cal O}_2}  + 
(\Delta M_s)_{{\cal O}_4} \,,
\end{eqnarray}
where
\begin{eqnarray}
(\Delta M_{B_s})_{{\cal O}_2}  &=&  2  \frac{1}{2 m_{B_s}} 
 \frac{m_b^2}{4 v^2} \left (\frac{1}{m_H^2} - \frac{1}{m_A^2} \right ) \eta_2(\mu_b) 
 \left ( -\frac{5}{12} \left (\frac{m_{B_s}}{m_b(\mu_b) + m_s(\mu_b)}\right )^2  m_{B_s}^2 f_{B_s}^2 \right ) 
 B_2(\mu_b)\, , 
 \nonumber \\
 \\
(\Delta M_{B_s})_{{\cal O}_4}  &=&  2  \frac{1}{2 m_{B_s}} 
 \frac{m_b m_s}{4 v^2} \left (\frac{1}{m_H^2} + \frac{1}{m_A^2} \right )  \eta_4(\mu_b) 
 \left (\frac{1}{2} \left (\frac{m_{B_s}}{m_b(\mu_b) + m_s(\mu_b)}\right )^2  m_{B_s}^2 f_{B_s}^2 \right )
 B_4(\mu_b) \,.
 \nonumber \\
\end{eqnarray}
For the bag model parameters we use the results from Ref.~\cite{becirevic} 
estimated in the quenched approximation on the lattice 
($\mu_b = m_b^{\rm RI-MOM} = 4.6$ GeV):
\begin{eqnarray}
B_2(\mu_b) = 0.82\,, \qquad B_4(\mu_b) = 1.16\,,
\end{eqnarray}
and the running of Wilson coefficients given at the same 
scale~\cite{becirevic}: $\eta_2(\mu_b) \simeq 2.03$, $\eta_4(\mu_b) 
\simeq 3.23$. The masses 
of the $b$ quark and the $s$ quark at the TeV scale are $m_b = 2.4$ GeV 
and $m_s = 45$ MeV, respectively. 

The current experimental value of $\Delta m_{B_s}$ is $116.4 \pm 0.5 
\times 10^{-10}$ MeV~\cite{pdg}, which agrees with the SM prediction to 
about 10\%.  Hence we limit our new physics contribution to $11.6 \times 
10^{-10}$ MeV.  
The allowed range for the masses of $H$ and $A$ Higgs bosons in the region 
greater than 1 TeV is shown in Fig.~1. Smaller masses, although possible, 
require fine tunning to satisfy the experimental $B_s- \bar{B}_s$ bound.
\begin{figure}
\begin{center}
\includegraphics[width=3in]{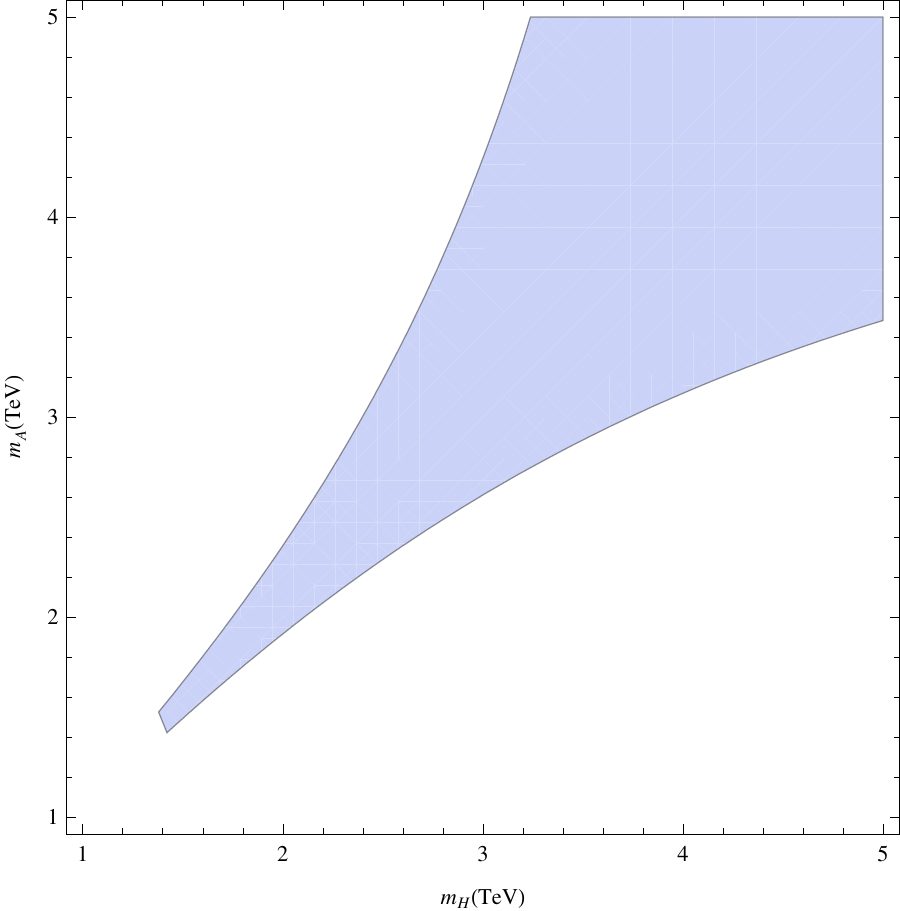} 
\end{center}
\caption{$B_s- \bar{B}_s$ mixing constraint for the masses $m_H$ and 
$m_A > 1 $ TeV. Masses less than 1 TeV require fine tunning to satisfy 
the constraint.}
\end{figure}

Note that in this approximation, there is no 
contribution to $b \to s \gamma$ from the new 
scalars, because of a residual $Z_2$ discrete symmetry under which 
$(c,s)$ are even, whereas $(H^\pm, H^0, A)$ and $(t,b)$ are odd. 

\section{The $3 \times 3$ Quark Mixing Matrix}

The introduction of the third Higgs doublet $\Phi_3$ will allow the $u$ and 
$d$ quarks to obtain mass, as well as mixing with the other quarks.  Since 
these are all small, it is natural to assume that $v_3$ is small.  (This 
also means that the $\Phi_3$ mass may be naturally large, as shown later.) 
In that case, the $h^0$ of Eq.~(12) is still a very good approximation to 
the one physical Higgs boson $h^0_{SM}$ of the SM, i.e. we have an 
explanation of why our specific three-Higgs doublet model has a 
mass eigenstate $h^0$ which is very close to $h^0_{SM}$.

The $3 \times 3$ quark mass matrices are given by~\cite{cfm04}
\begin{equation}
{\cal M}_d = \pmatrix{g_3^d v_3 & g_4^d v_3 & 0 \cr 
0 & g_1^d v_2 & -g_2^d v_2 \cr 0 & g_1^d v_1 & g_2^d v_1},
\end{equation}
and
\begin{equation}
{\cal M}_u = \pmatrix{g_3^u v_3^* & g_4^u v_3^* & 0 \cr 
0 & g_1^u v_1^* & -g_2^u v_1^* \cr 0 & g_1^u v_2^* & g_2^u v_2^*}.
\end{equation}
Note that $v_1,v_2$ in ${\cal M}_d$ are replaced by $v_2^*,v_1^*$ in 
${\cal M}_u$.  This means that for $v_1 \neq v_2$, there will be a 
mismatch in the $s-b$ and $c-t$ sectors.  Since $m_s << m_b$ and 
$m_c << m_t$, these mass matrices are simply diagonalized on the left: 
${\cal M}_d$ by
\begin{equation}
V_d = \pmatrix{1 & 0 & 0 \cr 0 & c' & -s' \cr 0 & s' & c'} 
\pmatrix{c_d & -s_d & 0 \cr s_d & c_d & 0 \cr 0 & 0 & 1},
\end{equation}
where $s'/c' = v_2/v_1$, and ${\cal M}_u$ by
\begin{equation}
V_u = \pmatrix{1 & 0 & 0 \cr 0 & s' & -c' \cr 0 & c' & s'} 
\pmatrix{c_u & -s_u e^{i \delta} & 0 \cr s_u e^{-i \delta} & c_u & 0 \cr 0 & 0 & 1}.
\end{equation}
We have rephased $d_R,s_R,b_R$, $u_R,c_R.t_R$ as well as $(u,d)_L$, 
$(c,s)_L$ so that only one complex phase $\delta$ remains.  Hence
\begin{eqnarray}
V_{CKM} = V_u^\dagger V_d  
&=& \pmatrix{c_u & s_u e^{i \delta} & 0 \cr -s_u e^{-i \delta} & c_u & 0 
\cr 0 & 0 & 1} \pmatrix{1 & 0 & 0 \cr 0 & c'' & s'' \cr 0 & -s'' & c''} 
\pmatrix{c_d & -s_d & 0 \cr s_d & c_d & 0 \cr 0 & 0 & 1} \nonumber \\ 
&=& \pmatrix{c_u c_d + c'' s_u s_d e^{i \delta} & -c_u s_d 
+ c'' s_u c_d e^{i \delta} & s'' s_u e^{i \delta} \cr -s_u c_d e^{-i \delta} 
+ c'' c_u s_d & s_u s_d e^{-i \delta} + c'' c_u c_d & s'' c_u \cr 
-s'' s_d & -s'' c_d & c''},
\end{eqnarray}
where $s''/c'' = ({c'}^2 - {s'}^2)/2 s' c'$.  Using the 2012 Particle Data Group  values~\cite{pdg}, 
we obtain
\begin{equation}
s'' = 0.04135, ~~ s_u = 0.08489, ~~ s_d = 0.20983, ~~ 
\cos{\delta} = -5.47 \times 10^{-3},
\end{equation}
with the CP violating parameter
\begin{equation}
J = s_u c_u s_d c_d (s'')^2 c'' \sin{\delta} = 2.96 \times 10^{-5}\,.
\end{equation}

\section{The Complete Higgs Sector}

To allow for $v_1 \neq v_2$, the symmetry $\Phi_1 \leftrightarrow \Phi_2$ 
must be broken.  This may be accomplished by adding to $V_{12}$ of Eq.~(10) 
the term $\mu^2_3 (\Phi_1^\dagger \Phi_1 - \Phi_2^\dagger \Phi_2)$.  The 
coefficient $\mu_3^2$ may be chosen naturally small, because $\mu^2_3 = 0$ 
results in the extra $Z_2$ symmetry already discussed.  In addition, we 
impose a new $Z_2$ symmetry so that $\Phi_3$ and $(u,d)_L$ are odd and 
all other fields are even on $Z_2$.  The purpose of this symmetry is to forbid 
the quartic term $\Phi_3^\dagger (\Phi_1 \Phi_2^\dagger \Phi_1 
+ \Phi_2 \Phi_1^\dagger \Phi_2) + h.c.$ which is allowed by $S_3$ (the 
reason for this will become clear later).  However, we also allow this new 
$Z_2$ symmetry to be broken softly by the bilinear term $\mu_4^2 \Phi_3^\dagger 
(\Phi_1 + \Phi_2) + h.c.$, which preserves the $\Phi_1 \leftrightarrow \Phi_2$ 
interchange symmetry of $V_{12}$.  The complete scalar potential of this 
model is then
\begin{eqnarray}
V_{123} &=& V_{12} + \mu^2_3 (\Phi_1^\dagger \Phi_1 - \Phi_2^\dagger \Phi_2) 
+ m_3^2 \Phi_3^\dagger \Phi_3 + [\mu_4^2 \Phi_3^\dagger (\Phi_1 + \Phi_2) + h.c.] 
\nonumber \\ 
&+& {1 \over 2} \lambda_4 (\Phi_3^\dagger \Phi_3)^2 + 
\lambda_5 (\Phi_3^\dagger \Phi_3)(\Phi_1^\dagger \Phi_1 + 
\Phi_2^\dagger \Phi_2) + \lambda_6 \Phi_3^\dagger (\Phi_1^\dagger \Phi_1 + 
\Phi_2^\dagger \Phi_2) \Phi_3 \nonumber \\ 
&+& [\lambda_7 (\Phi_3^\dagger \Phi_1)(\Phi_3^\dagger \Phi_2) + h.c.]\,.
\end{eqnarray}

Consider first $v_1 \neq v_2$, this results in a deviation of $h^0$ from 
$h^0_{SM}$ given by
\begin{equation}
h^0 - h^0_{SM} \simeq {(\lambda_1 - \lambda_2 + \lambda_3)(v_1^2 - v_2^2) \over 
2 \mu_2^2} H^0\,,
\end{equation}
where $(v_1^2 - v_2^2)/4v^2 = 0.0207$ from $s'' = 0.04135$ of Eq.~(23). 

Consider now $v_3 \neq 0$.  Since $m_3^2 > 0$ is assumed large,
\begin{equation}
v_3 \simeq {-\mu^2_4 (v_1 + v_2) \over m_3^2 + (\lambda_5 + \lambda_6) 
(v_1^2 + v_2^2) + 2 \lambda_7 v_1 v_2} \simeq {-2v \mu^2_4  \over m_3^2},
\end{equation}
and the mixing of $\phi_{3R}$ with $\phi_{1R} + \phi_{2R}$ is $v_3/(2v)$. 
This results in
\begin{equation}
h^0 - h^0_{SM} \simeq {v_3 m_h^2 \over 2 v m_3^2} {\rm Re}(\phi_3^0).
\end{equation}
The $h^0$ of our model is thus naturally equal to $h^0_{SM}$ in the 
symmetry limit $v_1=v_2$ and $v_3=0$, and the deviation is naturally 
suppressed for realistic values of $v_3^2 << v_2^2 \simeq v_1^2$.
If the quartic scalar term forbidden by the new $Z_2$ symmetry exists, then 
$h^0 - h^0_{SM} \simeq (2v_3/v) {\rm Re}(\phi_3^0)$ would replace Eq.~(30). 
This means that $h^0$ exchange itself will contribute too much to 
$K^0 - \bar{K}^0$ mixing.  With the relation in Eq.~(30), this contribution is 
negligible.

\section{Constraint on $\Phi_3$}

The exchange of $\phi_3^0$ directly contributes to $\Delta M_K$. 
The relevant effective interaction is given by
\begin{equation}
{s_d^2 c_d^2 m_d m_s \over v_3^2 m_3^2} (\bar{d}_L s_R)(\bar{d}_R s_L).
\end{equation}

This operator is again ${\cal O}_4$ from Eq.~(20), 
with the appropriate exchange of the 
quarks and the contribution to $\Delta m_K$ is analogous to Eq.~(23), i.e.
\begin{eqnarray}
 (\Delta M_{K})_{{\cal O}_4}  &=&  2  \frac{1}{2 m_{K}} 
s_d^2 c_d^2 \frac{m_s m_d}{v_3^2} \frac{1}{m_{3}^2}  \eta_{4K}(\mu_K) 
 \left (\frac{1}{2} \left (\frac{m_{K}}{m_s(\mu_K) + m_d(\mu_K)}\right )^2  
m_{K}^2 f_{K}^2 \right )
 B_{4K}(\mu_K) \,,
 \nonumber \\
\end{eqnarray}
with $\mu_K =2$ GeV. The relevant parameters are taken from 
Ref.~\cite{buras2}: 
\begin{eqnarray} 
B_{4K}^{\rm eff}(\mu_K) = \frac{1}{2} \left ( \frac{m_{K}}{m_s(\mu_K) + m_d(\mu_K)}\right )^2   B_{4K}(\mu_K)  = 19.31 \,,
\end{eqnarray}
where $B_{4K}(\mu_K) = 1.03$ itself, and the running of the Wilson coefficient 
gives $\eta_{4K}(\mu_K) = 4.87$~\cite{ciuchini}. 

Assuming that this contribution is no more than 20\% of the experimental 
value $\Delta M_K = 3.483 \pm 0.006 \times 10^{-12}$ MeV ~\cite{pdg}, we 
obtain 
\begin{equation}
v_3 m_3 > 6.0 \times 10^{4}~{\rm GeV}^2\,.
\end{equation}
If $v_3 = 10$ GeV, then $m_3 > 6$ TeV. 

So far, we have been able to show that $h^0$ is very close to $h^0_{SM}$, 
and that the other physical scalars are of order 1 to 10 TeV, from 
$B_s - \bar{B}_s$ and $K^0 - \bar{K}^0$ mixing respectively.  All other 
effective flavor-changing neutral-current interactions  in the quark sector 
such as $b \to s \gamma$ are suppressed.

\section{Specific Prediction}

Since the scalars of this model also have leptonic interactions, there 
are some lepton flavor violating processes in this model which are negligible in the 
SM.  The corresponding Lagrangian to Eq.~(17) for the 
$\mu - \tau$ sector is given by
\begin{eqnarray}
{\cal L}_Y &=& {m_\mu \over \sqrt{2} v} \left[H^+ \bar{\nu}_{\tau L} + 
\left( {H^0 + i A \over \sqrt{2}} \right) \bar{\tau}_L \right] \mu_R 
\nonumber \\ 
&+& {m_\tau \over \sqrt{2} v} \left[ H^+ \bar{\nu}_{\mu L} + 
\left( {H^0 + i A \over \sqrt{2}} \right) \bar{\mu}_L \right] \tau_R 
+ h.c.
\end{eqnarray}
Hence $b \to s \tau^- \mu^+$  proceeds again through the exchange of 
$H^0 + iA$ 
(but $b \to s \tau^+ \mu^-$ is suppressed by $m_\mu^2/m_\tau^2$). 

Experimentally the most interesting decay would be $B_s \to \tau^+ \mu^-$. 
The branching ratio for this decay can be written as 
\begin{eqnarray}
BR(B_s \to \tau^+ \mu^-) = \frac{m_{B_s}^5 f_{B_s}^2}{64 \pi} \tau_{B_s} m_{\tau}^2 
\left ( 1 - \frac{m_\tau}{m_{B_s}} \right )^2 \frac{1}{v^4} \left ( \frac{1}{m_H^2} + \frac{1}{m_A^2} \right )^2
\end{eqnarray}
We see in Fig.~2 
 that the branching ratio for this decay can be as high as $10^{-7}$ for masses 
of $H$ and $A$ bosons in the TeV range. 
\begin{figure}
\begin{center}
\includegraphics[width=4in]{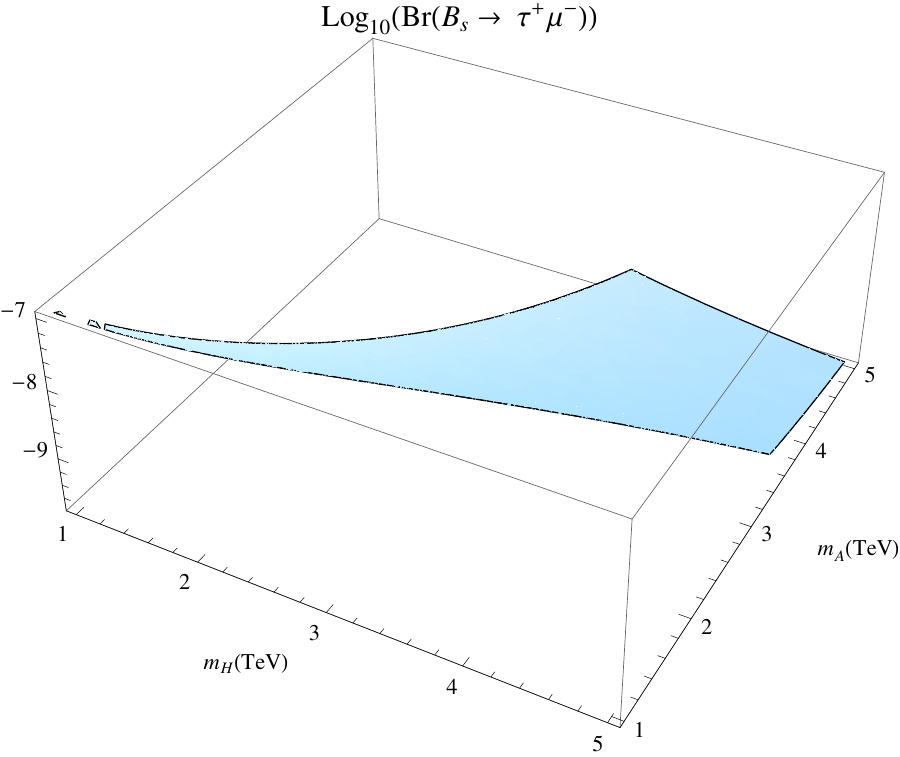}  
\end{center}
\caption{Prediction for $B_s \to \tau^+ \mu^-$ lepton flavor violation process in our model, with the masses of 
$H$ and $A$ bosons constrained by the $B_s-\overline{B}_s$ mixing from Fig.1.}
\end{figure}
This is a possible unique signature of this model.
For the less realistic masses smaller than 1 TeV, the $B_s \to \tau^+ \mu^-$ 
branching ratio can come up to $O(10^{-6})$.

\section{Conclusion}

In the post-Higgs era, any extension of the SM has to face the 
question of why it contains a light neutral scalar boson $h^0$ so much like to 
$h^0_{SM}$, in addition to being consistent with a myriad of phenomenological 
precision measurements.  We show how this is possible in a model~\cite{cfm04} 
proposed in 2004 based on the non-Abelian discrete symmetry $S_3$.  It has 
three Higgs doublets, and yet one becomes almost exactly that of the SM because of 
two approximate residual $Z_2$ symmetries.  It also explains 
why the quark mixing matrix is nearly diagonal, with just enough parameters 
to fit the data precisely.  The model is constrained principally by $B_s - \bar{B}_s$ 
and $K^0 - \bar{K}^0$ mixing, and has the unique prediction of $b \to s \tau^- 
\mu^+$ with a branching fraction for $B_s \to \tau^+ \mu^-$ decay as large as $10^{-7}$, but a strong 
suppression by the factor $m_\mu^2/m_\tau^2$ for $b \to s \tau^+ \mu^-$ decay.

\noindent \underline{Acknowledgment}~:~ This work is supported in part 
by the U.~S.~Department of Energy under Grant No.~DE-AC02-06CH11357. 
BM acknowledges support of the Fulbright foundation. 

\bibliographystyle{unsrt}

\end{document}